\documentclass[a4paper,11pt]{article} 
\usepackage[T2A]{fontenc}			
\usepackage[utf8]{inputenc}			
\usepackage{graphicx}
\graphicspath{}
\DeclareGraphicsExtensions{.pdf,.png,.jpg}
\usepackage{amsmath,amsfonts,amssymb,amsthm,mathtools,verbatim,alltt} 
\usepackage{braket}
\usepackage{wasysym}
\usepackage{cancel}
\usepackage{setspace}
\usepackage{slashed}
\usepackage{cite} 
\usepackage{authblk}
\everymath{\displaystyle}
\usepackage{mathrsfs}

\usepackage[colorlinks = true,
linkcolor = blue,
urlcolor  = blue,
citecolor = blue,
anchorcolor = blue]{hyperref}

\date{}
\title{Chiral effects: new trends}
\author[1,2]{G. Yu. Prokhorov}
\author[1,2]{O. V. Teryaev}
\author[2,1,3]{V. I. Zakharov}
\affil[1]{Joint Institute for Nuclear Research, Joliot-Curie str. 6, Dubna 141980, Russia}
\affil[2]{NRC Kurchatov Institute, Moscow, Russia}
\affil[3]{Pacific Quantum Center, 
Far Eastern Federal University, 10 Ajax Bay, Russky Island, Vladivostok 690950, Russia}

\begin{document}

\maketitle

\begin{center}
\section*{Abstract}
By chiral effects one  understands manifestations of chiral gauge anomaly and of gravitational chiral
anomaly in hydrodynamics. In recent two-three years our understanding of the chiral effects 
has considerably changed. Here we present mini-review of two topics. First, shift in understanding
symmetry which underlies the chiral magnetic effect and, second, interpretation of the chiral
kinematical effect uncovered recently.
\end{center}

\section{Introduction}
\label{sec intro}
Chiral effects have attracted a lot of attention in the last ten-fifteen years,
see, e.g., \cite{volume,kharzeev}.
The beauty of these effects is that they represent a quantum loop effects
in hydrodynamics which traditionally treated as belonging to classical physics (with the
exception of theory of superfluidity which is based on a particular mechanism and
this mechanism does not apply to chiral effects).

In particular, the gauge anomaly takes the form:
\begin{equation}\label{gauge}
\partial_{\alpha}J^{\alpha}_{5} = e^2C_5^{el}\cdot\vec{E}\cdot \vec{B}~,
\end{equation}
where $J^5_{\alpha}$ is axial-vector current, $C_5^{el}$ is a constant which depends 
on the choice of fundamental constituents, $\vec{E}$ and $\vec{B}$
are electric and magnetic fields, respectively, $e$ is electromagnetic coupling. 
By the chiral magnetic effect (CME) one understands
the electric current flowing in the direction of the magnetic field:
\begin{equation}\label{cme}
\vec{J}^{el}~=~e^2 C^{el}_5 \cdot \mu_5 \vec{B}~,
\end{equation}
where $\mu_5$ is the axial chemical potential and $C^{el}_5$ is the same constant 
which enters definition of the anomaly (\ref{gauge}) .
As indicated by the
$e^2$ factor, (\ref{cme}) looks as a Hall-type current.
The guess turns true once one takes into account
the specific hydrodynamic interaction $\hat{H}_{hydro}$:
\begin{equation}\label{effective}
 \hat{H}_{hydro}~=~ \hat{H}_{field ~theory}-
\mu_{el}~ \hat{Q}_{el}~-~\mu_5~\hat{Q}_5~,
\end{equation}
where $\hat{Q}_{el}$ and $\hat{Q}_5$ are electric and axial charges.
The use of (\ref{effective}) assumes that the charges are conserved.

In presence of external gravitational field divergence of the axial current $J_5^{\alpha}$ receives
a further contribution:
\begin{equation}\label{gravitational}
\nabla_{\alpha}J^{\alpha}_5~=~C_5^{grav}R\tilde{R} ~, 
\end{equation}
where $\nabla
_{\alpha}$ is the covariant derivative, 
$R\tilde{R}=(1/2)\epsilon^{\mu\nu\alpha\beta}R_{\alpha\beta}^{\gamma\delta}
R_{\gamma\delta\mu\nu}$, $R_{\mu\nu\alpha\beta}$ is the Riemann tensor, 
$C_5^{grav}$ is a constant which depends on spin of fermions and is known.
No generalization of Eq. (\ref{effective}) to the case of gravitational field is known.
Hence, no analog of (\ref{cme}) is known either. For recent development see Sect. III.

\section{From chiral symmetry to diffeomorphism}
In this section we describe briefly recent progress in understanding physics behind the chiral
magnetic effect.

Let us first note  that  using the well known analogy between  the magnetic field  $\vec{B}$ 
and angular velocity $\vec{\Omega}$
generates so called  chiral vortical effect (CVE) from Eq. (\ref{cme}):
\begin{equation}\label{specific}
 \vec{J}_{5,CVE}~=~\mu^2_{el}C_5^{el}\vec{\Omega},
\end{equation}
 Let us emphasize that this current exists in absence of $\vec{E}, \vec{B}$ and, therefore,
is to be conserved. Nevertheless, it is proportional to the constant in front of the gauge anomaly, 
$C_5^{el}$. The paradox is strengthened by observation that there is 
  no place for a new Noether current, since there is no free symmetry to associate with 
such a current. A possible way out is to assume that the
conservation of the current (\ref{specific}) is specific for absence of dissipation, 
or for ideal fluid, see \cite{avdoshkin,zakharov}.
And, indeed, one can demonstrate that the current (\ref{specific}) is conserved in case of ideal fluid.

For ideal fluid, however, there exists another
conserved axial charge (not chiral!), see, e.g., \cite{jackiv} :
\begin{equation}
 Q_{helicity}~=~(const)\int d^3x~ \vec{v}\cdot (\vec{\nabla} \times
\vec{v} )~~,
\end{equation}
or
\begin{equation}\label{helicity}
~J^{\alpha}_{helicity}~\sim~
\epsilon^{\alpha\beta\gamma\delta}u_{\beta}\partial_{\gamma}u_{\delta}~,
\end{equation}
~where~$u_{\alpha}$~is the fluid~4-velocity.

 The crucial point is that the current (\ref{helicity}) is indeed conserved but rather
as a consequence of {\it diffeomorphism} symmetry of ideal fluid. 
 Thus, chiral symmetry is apparently  embedded into diffeomorphism and
we encounter a
fundamental change of symmetry. Thus, let us pause here and make a few remarks. 
 
First of all, the conserved helicity current (\ref{helicity}) is not proportional to the electromagnetic
coupling and represents a classical current while the commonly discussed CME current
(\ref{cme}) is a quantum, or loop effect.
To generate a  CME current and its divergence  back from (\ref{helicity}) 
replace $\partial_{\alpha} ~\to~\nabla_{\alpha}$. Then we come to
amusing {non-relativistic} analogs to chiral anomalies \cite{abanov,abanov1,mitra}:
\begin{eqnarray}\label{anomalies}
\partial_{\alpha}J^{\alpha}_5~\sim~\vec{E}\cdot\vec{B}~~~~(as ~``usual")~,\\ \nonumber
\partial_{\alpha}J^{\alpha}_{el}~\sim~\vec{E}_5\cdot\vec{B}\equiv~
\vec{\nabla}\mu_{5}\cdot\vec{B}~,
\end{eqnarray}
where $\vec{E} \equiv \vec{\nabla}\mu$, $\vec{E}_5\equiv \vec{\nabla}\mu_5$.
In other words, we generate anomaly in electromagnetic current. And this seems to be a 
fundamental challenge to the theory.  But actually it is not: the current is not conserved 
simply because
in non-inertial frames the 4-volume depends on the choice of the frame, and this factor 
is to be taken into account \cite{mitkin}.

Here we summarize our findings in the current section:
\begin{itemize}
\item{The newly discovered anomalies (\ref{anomalies}) are not 
related to any ultraviolet divergences. }
\item{Since we started with a non-relativistic current (\ref{helicity})
the anomalies (\ref{anomalies}) are non-relativistic in nature as well.
The non-relativistic anomalies match the commonly known UV anomalies only up to a factor.}
\item{Non-conservation of density of electromagnetic current is found.}
\end{itemize}
The overall conclusion to this section is the conjecture that
chiral fluid of massless quarks at short distances
 might look as an ideal non-relativistic fluid at large distances.
 The reason for the conjecture is that there is matching of the corresponding anomalies
(variation of the  't Hooft consistency condition).

\section{Gravitational anomaly and hydrodynamics}

Let us start this section with a very brief reminder of some of the 
basics of the General Relativity (GR).
General Relativity is built on metric tensor
\begin{equation}
g_{\mu\nu}~\approx~\eta_{\mu\nu}~+~h_{\mu\nu}~.
\end{equation}
Gauge transformations, analog of 
$\delta{A_{\mu}=\partial_{\mu}\Lambda}$, look as:
\begin{equation}
\delta{h_{\mu\nu}}~=~\partial_{\mu}\epsilon_{\nu}+
\partial_{\nu}\epsilon_{\mu}~,
\end{equation}
where $\epsilon_{\mu}$ are arbitrary (smooth) functions.
Gauge invariant field, analog of $\vec{E}, \vec{B}$,  is Riemann tensor 
$R_{\alpha\beta\gamma\delta}$
which involves $(\partial_{\alpha}\partial_{\beta} h_{\mu\nu})$ ~and~
$(\partial_{\alpha}h_{\mu\nu})^2$.
Gravitational anomaly is built on the curvature $R_{\alpha\beta\gamma\delta}$,
see Eq. (\ref{gravitational}).

A new question, not much explored so far is:
``In absence of sizable curvature, 
what is left of the anomaly?''.
Without being rigorous, let us try to give a preliminary answer:
what is left is the equivalence principle. It states that the manifestations of a gravitational field
$h_{\mu\nu}$ and of going into a non-inertial frame are identical to each other. Both are gauge
dependent but their identity is a gauge invariant statement. In a bit more formal language, 
the covariant derivatives are also gauge invariant, not only curvature.

Two basic non-inertial frames, considered already by Einstein are
{accelerated ~and~rotated ~frames}  
(corresponding $h_{\mu\nu}$ are easy to find out in textbooks).
In applications to a liquid, notably quark-gluon plasma, acceleration and rotation 4-vectors 
are given by:
\begin{equation}
a_{\mu}~=~u^{\alpha}\partial_{\alpha}u_{\mu},~~~
\omega_{\mu}~=~\frac{1}{2}\epsilon_{\mu\nu\rho\sigma}u^{\nu}\partial^{\rho}u^{\sigma}~~
~(u^{\nu}~is~4-velocity)~.
\end{equation}
In particular, in absence of curvature  the axial current  is
still not vanishing and fixed uniquely in terms of kinematical
quantities $a_{\mu}, w_{\mu}$.
In case of spin 1/2:
\begin{equation}
J_5^{\alpha,kinematic}~=~-\frac{1}{24\pi^2}\Big(3 a_{\mu}^2+w_{\mu}^2\Big)
w^{\alpha}~.
\end{equation}
Moreover, a particular combination of coefficients in front of $(a_{\mu})^2=a_{\mu}a^{\mu}$ and $(w_{\mu})^2=w_{\mu}w^{\mu}$
can be expressed in terms of the coefficient  in front of the anomaly 
$C_5^{grav}$ \cite{prokhorov1}.
This relation  is called kinematical vortical effect effect (KVE) \cite{prokhorov1}.

The idea and  details of derivation of KVE are summarized in other talks at this conference.
Here,
we add  interpretation of the KVE. The key point is the
Unruh effect.
According to Unruh \cite{unruh},
observer moving with acceleration $a$ with respect to Minkowskian vacuum
 sees thermal distribution of particles with  temperature
\begin{equation}
T_{Unruh}~=~\frac{a}{2\pi}~,
\end{equation}
while an observer at rest sees no particles.
For many, it might sound disappointing that the Unruh effect is a kind of observer-dependent.
In fact, it is dynamical. Indeed,  by virtue of the equivalence principle, accelerated frame is
equivalent to vacuum placed in (strong) gravitational field resulting in the same 
acceleration
$a$. Naturally, such a field produces  particles and is obviously dynamical.

After these remarks, we can go directly to the conclusion to the current section.
  Kinematical effects  refer to non-inertial frames, describing
results of measurements on the Unruh sample of particles.
 The validity of this explanation can be substantiated by a direct analysis
of derivation of KVE in \cite{prokhorov1}. There is a subtle point:
 rotation does not produce work. For this reason the ``Unruh sample of particles''
is determined by acceleration alone, not by rotation.

\section{Duality of statistical and gravitational approaches}

Properties of fluids in equilibrium are evaluated statistically 
in terms  of density operator, or
effective interaction: 
 \begin{equation} \label{density}
\hat{H}_{eff} ~=~
-\vec{\Omega}\cdot \hat{\vec{M}}-\vec{a}\cdot\hat{\vec{K}}~,
\end{equation}
\vspace{0.1cm}
where $\hat{\vec{M}}$ is operator of  angular momentum and $\hat{\vec{K}}$ is 
operator of boost
(for details see, e.g. \cite{becattini}).

In field theory, gravitational interaction is described by fundamental interaction Lagrangian:
\begin{equation}
{\delta L}_{fund}~=~-\frac{1}{2}\hat{\Theta}^{\alpha\beta}h_{\alpha\beta}~,
\end{equation}
where $\Theta^{\alpha\beta}$ is the energy-momentum tensor of matter, $h_{\alpha\beta}$ is
the gravitational potential accommodating the same $\vec{\Omega}$, $\vec{a}$
as the density operator (\ref{density}).
Furthermore, one
evaluates ``external probes'',  $<\Theta^{\alpha\beta}>, <J^{\alpha}_5>$ 
within both approaches, statistical and gravitational.
For the same values of $a_{\mu}, w_{\mu}$, the expectation is
that  results turn to be the same. The expectations are confirmed on many examples 
\cite{prokhorov2}. 
No surprise, since our example does satisfy a standard  set of conditions
to be satisfied for two theories to be dual to each other:
\begin{itemize}
\item{Two theories are to have the same symmetry pattern. And, indeed,  both ideal 
fluid and gravity are diffeomorphic invariant.}
\item{Infrared vs ultraviolet sensitivity. Statistics is not valid at short distances, while
field theory does need UV regularization. Thus, the two approaches are complementary to each other.}
\item{Both theories allow to evaluate a common set of observables.}
\end{itemize}
 As a result, evaluation of the kinematical effect can be rewritten as 
regularization, via acceleration, of gravitational chiral anomaly. Such a regularization
is infrared-sensitive, unlike the standard derivation of the anomalous divergence of the current,
which is UV sensitive. To elaborate, how general  this observation is and weather it could be developed
into a novel type of  regularization, is a task for the future.

\end{document}